\documentclass[12pt,prb,aps,preprint]{revtex4}

\usepackage{graphics}
\usepackage{graphicx}
\usepackage{amsfonts}
\usepackage{amsmath}
\usepackage{float}
\usepackage{epsfig}
\usepackage{subfigure}

\begin{document}
\title{Nonequilibrium static growing length scales in supercooled liquids on approaching 
the glass transition}

\author{{\' E}tienne Marcotte}
\affiliation{Department of Physics, Princeton University, Princeton, New
Jersey 08544, USA}
\author{Frank H. Stillinger}
\affiliation{Department of Chemistry, Princeton University, Princeton, New
Jersey 08544, USA}
\author{Salvatore Torquato}
\affiliation{Department of Chemistry, Princeton University, Princeton, New
Jersey 08544, USA}
\affiliation{Princeton Institute for the Science and Technology of Materials,
Princeton University, Princeton, New Jersey 08544, USA}
\affiliation{Department of Physics, Princeton University, Princeton, New Jersey
08544, USA}
\affiliation{Program in Applied and Computational Mathematics, Princeton
University, Princeton, New Jersey 08544, USA}

\begin{abstract}
 The small wavenumber $k$ behavior of the structure factor $S(k)$ of overcompressed amorphous hard-sphere
configurations was previously studied for a wide range of densities up to the maximally random jammed state, which can
be viewed as a prototypical glassy state [A. Hopkins, F. H. Stillinger
and S. Torquato, Phys. Rev. E, 86, 021505 (2012)]. It was found that a precursor to the glassy jammed state 
was evident  long before the jamming density was reached as measured by a
growing {\it nonequilibrium} length scale  extracted from the volume integral of the direct 
correlation function $c(r)$, which becomes long-ranged as the 
critical jammed state is reached. The present study extends that work by investigating
via computer simulations two different
atomic models: the  single-component Z2 Dzugutov potential in three dimensions
and the binary-mixture Kob-Andersen potential in two dimensions. Consistent
with the aforementioned hard-sphere study, we
demonstrate that for both models a signature of the glass transition 
is apparent well before the transition temperature  is reached 
as measured by the length scale determined from 
from the volume integral of the direct correlation function in the
single-component case and a generalized direct correlation function in the
binary-mixture case.
The latter quantity is obtained from a generalized Orstein-Zernike integral
equation for a certain decoration of the atomic point configuration.
We also show that these growing
length scales, which are a consequence of the long-range nature of the
direct correlation functions, are intrinsically nonequilibrium in nature
as determined by an index $X$ that is a measure of deviation from thermal
equilibrium.
It is also demonstrated that this nonequilibrium index, which increases
upon supercooling, is correlated with a characteristic relaxation time scale.
\end{abstract}

\maketitle
\section{Introduction}

A quantitative understanding of nature of the physics of the glass transition is one of the
most fascinating and challenging problems in materials science and condensed-matter
physics. A sufficiently rapid quench of a liquid from above its freezing temperature into
a supercooled regime can avoid crystal nucleation to produce a glass with a
relaxation time that is much larger than experimental time scales, resulting in
an amorphous characteristic state (without long-range order) that is simultaneously rigid
\cite{CLPCMP1995}. A question that has received considerable attention 
in recent years is whether the growing relaxation times
under supercooling have accompanying growing structural length scales.
Two distinct schools of thought have emerged to address this question.
One asserts that static structure of a glass, as measured by pair correlations, is indistinguishable
from that of the corresponding liquid. Thus, since there is no signature of increasing static
correlation length scales accompanying the glass transition, it identifies
growing dynamical length scales. \cite{BBBKMR2007a,KDS2009a,CG2010a}
The other camp contends that there is a static growing length scale of {\it thermodynamic} origin
\cite{LW2006a,HMR2012a} and therefore one need not look for growing length
scales associated with the dynamics. 

In the present paper, we employ both theoretical and computational methods to study 
two different atomic glass-forming liquid models that support an alternative view, 
namely, the existence of a growing {\it static} length scale as the temperature
of the supercooled liquid is decreased that is intrinsically {\it nonequilibrium} in
nature. This investigation extends recent previous work \cite{Ho12} in which this conclusion
was first reached by examining overcompressed hard-sphere liquids
up to the maximally random jammed (MRJ) state. \cite{TTD2000a}
(For a hard-sphere system, compression qualitatively plays the
  same role as decreasing the temperature in an atomic or molecular system; see
  Ref. \onlinecite{TS2010a}.) The MRJ state under the strict-jamming constraint is a
{\it prototypical} glass in that it lacks any long-range order but is perfectly
rigid such that the elastic moduli are unbounded.\cite{TS2007a,TS2010a} This
endows such packings with the special hyperuniformity attribute.
A statistically homogeneous and isotropic single-component
point configuration at 
number density $\rho$ is hyperuniform if its structure factor 
\begin{equation}
S(k) = 1 +\rho {\widetilde h}(k)
\label{Sk}
\end{equation}
tends to zero as the wavenumber $k \rightarrow 0$,\cite{TS2003a} where
$h(r) \equiv g_2(r) -1$ is the total correlation function, $g_2(r)$ is the 
pair correlation function, and $\widetilde{h}(k)$ is the Fourier transform of 
$h(r)$. This condition implies that infinite-wavelength density
fluctuations vanish. 

It was theoretically established that hyperuniform point distributions are at an
``inverted'' critical point in that the direct correlation function $c(r)$, rather than the total
correlation function $h(r)$, becomes long-ranged, i.e., it decays more slowly than $-1/r^d$ in $d$-dimensional
Euclidean space $\mathbb{R}^d$, where $r$ is the radial distance. \cite{TS2003a} The Fourier transform of
direct correlation function ${\widetilde c}(k)$ is defined via the Ornstein-Zernike equation: \cite{HMTSL2006}
\begin{equation}
{\widetilde c}(k) = \frac{{\widetilde h}(k)}{1+\rho {\widetilde h}(k)}=\frac{S(k) - 1}{\rho S(k)}.
\label{Ck}
\end{equation}
It is immediately clear from this definition that the real-space volume integral
of the direct correlation function $c(r)$ diverges to minus infinity for any
hyperuniform system, since the denominator of (\ref{Ck}) tends to zero, i.e.,
\begin{equation}
\lim_{k \rightarrow 0} {\widetilde c}(k) = \int_{{\mathbb R}^d} c(r) d{\bf r} \rightarrow -\infty
\end{equation}

MRJ packings of identical spheres possess a special type of hyperuniformity 
such that $S(k)$ tends to zero linearly in $k$ as $k \rightarrow 0$,
implying quasi-long-ranged negative pair correlations (anticorrelations)
in which $h(r)$ decays as a power law $-1/r^4$
or, equivalently,  a direct correlation function that decays as
$c(r) \sim -1/r^2$ for large r, as dictated by Eq. (2).\cite{DST2005a} 
These anticorrelations reflect an unusual spatial patterning of regions of
lower and higher local particle densities relative to the system density.
This quasi-long-range behavior of $h(r)$ 
is distinctly different from typical liquids in
equilibrium, which tend to exhibit more rapidly decaying pair correlations,
including exponential decays.

Reference \onlinecite{Ho12} examined  overcompressed hard-sphere
configurations that follow Newtonian dynamics for a wide range of densities 
up to the MRJ state. 
A central result of this study was to establish that a precursor to the glassy 
jammed state was evident long before the MRJ density was reached
as measured by an associated growing length scale, extracted
from the volume integral of the direct correlation function $c(r)$, which of course
diverges at the ``critical'' hyperuniform MRJ state. 
It was also shown  that the nonequilibrium signature of 
the aforementioned quasi-long-range anticorrelations, which was quantified
via a {\it nonequilibrium index} $X$, emerges well before the jammed state was
reached. 

These results for nonequilibrium amorphous hard-sphere packings suggest that the direct correlation function of 
supercooled atomic models in which the atoms possess both repulsive and
attractive interactions should provide a robust {\it nonequilibrium} static growing length scale as the temperature is decreased to the glass transition and
below. Here we show that this is indeed the case
by extracting length scales associated with standard and generalized
direct correlation functions. In particular, we study
the single-component Z2 Dzugutov potential in three dimensions
and the binary-mixture Kob-Andersen potential in two dimensions. 
The Z2 Dzugutov potential for a single-component many-particle system 
in three dimensions  has the following form:\cite{Do03}
\begin{equation}
v(r) = \left\{ \begin{array}{ll}
a \frac{\displaystyle e^{\eta r}}{\displaystyle r^3} \cos(2 k_f r) + b \left(\frac{\displaystyle \sigma}{\displaystyle r}\right)^n + V_0, & \quad r < r_c , \\
0, & \quad r \ge r_c .
\end{array} \right.
\label{DZ}
\end{equation}
The first term in (\ref{DZ}) models Friedel oscillations for a metal 
with Fermi wave vectors of magnitude $k_F$, while the second term adds a 
strong repulsion for sufficiently small interparticle separations.
The parameters $a$ and $b$ control the relative strengths of both
contributions and define the energy scale. The cutoff $r_c$ is selected to be
at the third minimum of the potential, while the constant $V_0$ is present
to make the potential
continuous at the cutoff. The parameters $\eta$, $\sigma$, and $n$
control the shapes of both functions in (\ref{DZ}).
The Kob-Andersen model for a two-dimensional binary mixture is given by a  truncated Lennard-Jones potential:\cite{Ko94}
\begin{equation}
v_{\alpha \beta}(r) = \left\{ \begin{array}{ll}
4 \epsilon_{\alpha \beta} \left[ \left(\displaystyle\frac{\sigma_{\alpha \beta}}{r}\right)^{12} - \left(\displaystyle\frac{\sigma_{\alpha \beta}}{r}\right)^{6} \right] + V_{0\alpha\beta}, & \quad r < 2.5 \sigma_{\alpha \beta} , \\
0, & \quad r \ge 2.5 \sigma_{\alpha \beta} .
\end{array} \right.
\label{KA}
\end{equation}
The parameter $\epsilon_{\alpha\beta}$ controls the strength of the attraction
between two particles of species $\alpha$ and $\beta$, while 
$\sigma_{\alpha\beta}$ is equal to $2^{-1/6}$ times the distance between
both particles at which the attraction is maximal.

It is known that overcompressing a hard-sphere system is analogous to 
supercooling a thermal liquid, but to what extent does this analogy hold?
Roughly speaking, a rapid densification of a monodisperse hard-sphere system
leads to the terminal MRJ state (with packing fraction of about 0.64),
which we have noted is a prototypical glass.\cite{TS2010a}
At this singular state, the system is never able to relax and hence the
associated relaxation time is infinite.\cite{Rintoul1996} Slower
densification rates lead to other jammed states with packing fractions
higher than 0.64.\cite{TS2010a} Moreover, it has been shown that below 0.64,
metastable hard-sphere systems have bounded characteristic relaxation
times,\cite{Rintoul1996,Schweizer2007} including the range of packing
fractions of about $0.58 \sim 0.60$ (depending on the densification rate)
that has been interpreted to be the onset
of a kinetic glass transition.\cite{Schweizer2007}
Above a particular hard-sphere glass-transition density, 
the system is able to
support a shear stress on time scales small compared to a
characteristic relaxation time. Clearly, increasing the density of a
hard-sphere system plays the same role as decreasing temperature of a
thermal liquid. In a thermal system, a glass at absolute zero
temperature has an infinite relaxation time classically, and hence this state
is the analog of the hard-sphere MRJ state. The glass transition
temperature $T_g$, which depends on the quenching rate and possesses a
bounded characteristic relaxation time, is analogous to the aforementioned
kinetic transition in hard spheres. These strong analogies between glassy
hard-sphere states and glassy atomic systems lead one to believe that the
results of Ref.~\onlinecite{Ho12} for the former extend to the latter.
Indeed, here we demonstrate that the aforementioned length scales grow
as as the temperature is decreased to the glass transition $T_g$ and below.
Moreover, we show that the nonequilibrium index $X$, previously
shown\cite{Ho12} to increase as a hard-sphere system is densified to the
MRJ state, also grows for $T < 2.2 T_g$.
This nonequilibrium index is also shown to be correlated with an
early relaxation time $\tau$.

In Sec.~\ref{sec:struct_signatures}, we introduce two generalizations of
the direct correlation function $c(r)$ which apply for two-component
systems. In Sec.~\ref{sec:sim_details} we describe the numerical
techniques and parameters used in our simulations, while in 
Sec.~\ref{sec:results} we present the results we extract from these 
simulations. The latter includes the demonstration of the existence of
growing nonequilibrium static length scales upon supercooling the two
atomic-liquid models that we consider.
Moreover, we show that the nonequilibrium index $X$ is positively correlated
with an early relaxation time, both of which increase as the temperature
is decreased to the glass transition temperature and below.
We conclude in Sec.~\ref{sec:conclusion} with a summary
of our results and of their impact.

\section{Structural Signatures of Large-Wavelength Density Fluctuations in Binary Mixtures \label{sec:struct_signatures}}

It has been shown that for maximally random jammed binary sphere packings,
the standard structure factor $S(k)$, determined from the particle centroids,
cannot be used to ascertain whether the system is hyperuniform, unlike the
single-component MRJ sphere packing.\cite{ZJT2011a,endnote5}
Instead it was shown that the spectral density $\widetilde{\chi}(k)$,
defined below, can be employed to determine whether a binary MRJ packing
is hyperuniform, since it vanishes as $k \rightarrow 0$.
We will show below that one must modify the spectral density for particles
interacting with soft (non-hard-core) pair potentials because particle-shape
information is required in order to ascertain whether the system is
hyperuniform or nearly hyperuniform.
For particles interacting with a hard-core repulsion, the particle
shapes are obviously the hard cores, but for
non-hard-core interactions, such as in the Kob-Andersen model
studied in this paper, one must determine a self-consistent procedure to
assign particle shapes to each point particle. In addition, for such soft
binary mixtures, the standard direct correlation function $c(r)$, applicable
to monodisperse systems, must be generalized.

In this section, we present
two generalizations of $c(\mathbf{r})$ for polydisperse systems: one that is
based on the spectral density (Sec.~\ref{sec:single_component_case}), and
another that is based on the matrix version of the structure factor
(Sec.~\ref{sec:mixture_OZ}).

\subsection{Single-Component Ornstein-Zernike Equation}

For a statistically homogeneous and isotropic single component system,
the Ornstein-Zernike (OZ) equation\cite{HMTSL2006}
defines the direct correlation 
function $c(r)$ in term of the total correlation function 
$h(r) = g_2(r) - 1$ and the system point density $\rho$:
\begin{equation}
h(r) = c(r) + \rho h(r) \otimes c(r) ,\label{eqn:OZ_realspace}
\end{equation}
where $\otimes$ denotes a convolution.
After taking the Fourier transform of Eq.~(\ref{eqn:OZ_realspace}) and
introducing the structure factor 
$S(k) = 1 + \rho \widetilde{h}(k)$, we get
\begin{equation}
\widetilde{h}(k) = \widetilde{c}(k) + \rho \widetilde{h}(k) \widetilde{c}(k),
\end{equation}
which is equivalent to expression (\ref{Ck}).

As noted in the introduction, we see from relation (\ref{Ck}) that for 
hyperuniform single-component systems, i.e., 
$\lim_{k \rightarrow 0} S(k) = 0$, 
$\widetilde{c}(k)$ diverges toward $-\infty$ as $k$ approaches 0.

\subsection{Generalization of the Ornstein-Zernike Equation for ``Two-Phase'' Decorations\label{sec:two_phase_decoration}}

As indicated in the beginning of the section, we must obtain a modified
version of the direct correlation function for binary mixtures
in which the particles interact with non-hard-core pair potentials in
order to detect hyperuniformity or near-hyperuniformity.
This function must be defined to be as general as possible. In particular,
it must be equivalent to the usual direct correlation function in the
case of a single-component system. We shall therefore start by determining
what this modified function would be in the single-component case in order to
provide insight for the more general case of multiple-component systems.
This will be done by decorating the underlying point configuration with
nonoverlapping spheres. We first describe the single-component case and then
the mixture case.

\subsubsection{Single-Component Case\label{sec:single_component_case}}

Consider a configuration of $N$ points within a large volume $V$ in which the minimum pair separation is the distance $R$. 
Now let us decorate this configuration by circumscribing spheres of radius $R$ around each of the points, leading to a configuration of $N$ nonoverlapping spheres of radius $R$. In this case, the particle phase indicator $\mathcal{I}(\mathbf{x})$ in terms of the positions of the sphere
centers $\mathbf{r}_1,\mathbf{r}_2,\ldots, \mathbf{r}_N$
is:\cite{Torquato1985,TorquatoBook}
\begin{equation}
\mathcal{I}(\mathbf{x}) = \sum_{i=1}^N m(|\mathbf{x} - \mathbf{r}_i|; R) ,
\end{equation}
where $m(r;R)$ is the single-inclusion indicator
function given by
\begin{equation}
m(r;R) \equiv \Theta(R - r) =
\left\{ \begin{array}{ll}
1, & \quad r \le R, \\
0, & \quad r > R.
\end{array} \right.
\label{ind1}
\end{equation}
The two-point correlation function
$S_2(r)=\langle I({\bf x}) I({\bf x}+{\bf r}) \rangle$ for such a statistically homogeneous
and isotropic distribution of nonoverlapping spheres, equal to the probability of finding
two points, separated by the distance $r\equiv |{\bf r}|$, anywhere in the region occupied by the spheres, has been shown to
be given by the following sum of two terms:\cite{Torquato1985,TorquatoBook}
\begin{equation}
S_2(r) = \rho m(r) \otimes m(r) + \rho^2 m(r) \otimes g_2(r) \otimes m(r) \;,
\end{equation}
where $\rho=\lim_{V\rightarrow \infty} N/V$ is the number density and
angular brackets denote an ensemble average.
The quantity $\rho m \otimes m$ is the \emph{self-correlation} term,
which is equal to the
probability of finding two points inside the same sphere, and
$\rho^2 m \otimes g_2 \otimes m$ is the two-body correlation,
the probability of finding two points in two different
spheres.
The autocovariance function $\chi(r)$ is:
\begin{eqnarray}
\chi(r) & \equiv & S_2(r) - \rho^2 v_1^2 = \rho m(r) \otimes m(r) + \rho^2 m(r) \otimes g_2(r) \otimes m(r) - \rho^2 v_1^2 , \nonumber \\
 & = & \rho m(r) \otimes m(r) + \rho^2 m(r) \otimes h(r) \otimes m(r) , \label{eqn:chi_formula}
\end{eqnarray}
where 
\begin{equation}
v_1(R) = \int m(r; R) d\mathbf{r} = \frac{\pi^{\frac{d}{2}} R^d}{\Gamma(1 + \frac{d}{2})}
\end{equation}
is the volume of a $d$-dimensional 
sphere of radius $R$ [$v_1(R)=4\pi R^3/3$ for $d=3$ and $v_1(R)=\pi R^2$
for $d=2$]. Taking the Fourier transform of Eq.~(\ref{eqn:chi_formula}) yields
\begin{equation}
\widetilde{\chi}(k) = \rho \widetilde{m}^2(k) + \rho^2 \widetilde{m}^2(k) \widetilde{h}(k) = \rho \widetilde{m}^2(k) S(k) , \label{eqn:chi_from_S}
\end{equation}
where $S(k)$ is the structure factor define in (\ref{Sk}). 
One can see from this equation that if
the decorated ``two-phase'' nonoverlapping sphere system is hyperuniform,
both $S(k)$ and $\widetilde{\chi}(k)$ go to zero as $k \rightarrow 0$
(phase in this context does not refer to a thermodynamical phase,
but to either the particle or the void phase).

In order to manage the extension of the standard direct correlation function
that correspond to the autocovariance function $\chi(k)$, 
we present the following analysis.
The self-correlation term in relation (\ref{eqn:chi_formula}) must be
subtracted because in its present form $\chi(r)$ is not analogous to $h(r)$.
Thus, we introduce a modified autocovariance $H(r) = S_2(r) - \rho m(r) \otimes m(r)$, given explicitly by:
\begin{equation}
H(r) = \rho^2 m(r) \otimes h(r) \otimes m(r) . \label{eqn:xi_definition}
\end{equation}
Taking the Fourier transform of Eq.~(\ref{eqn:xi_definition}) leads to:
\begin{equation}
\widetilde{H}(k) = \rho^2 \widetilde{m}^2(k) \widetilde{h}(k) = \widetilde{\chi}(k) - \rho \widetilde{m}^2(k) .
\end{equation}
We can now define a new direct correlation function $C(r)$ using $H(r)$:
\begin{equation}
H(r) = C(r) + Q(r) \otimes C(r) \otimes H(r) , \label{eqn:raw_oz_equation}
\end{equation}
where $Q(r)$ is a function which is to be chosen
such that $\lim_{k \rightarrow 0} \widetilde{C}(k)$ diverges 
for any hyperuniform system, for which $\widetilde{\chi}(k) \rightarrow 0$
as $k \rightarrow 0$.
\begin{eqnarray}
\widetilde{H}(k) & = & \widetilde{C}(k) + \widetilde{Q}(k) \widetilde{C}(k) \widetilde{H}(k) , \label{eqn:fourier_oz_equation}
 \\
\widetilde{C}(k) & = & \frac{\widetilde{\chi}(k) - \rho \widetilde{m}^2(k)}{1 + \widetilde{Q}(k) \left(\widetilde{\chi}(k) - \rho \widetilde{m}^2(k)\right)} . \label{eqn:c_and_q}
\end{eqnarray}
For $\lim_{k \rightarrow 0}\widetilde{C}(k)$ to diverge for hyperuniform
systems, we require that the 
denominator of the right side of Eq.~(\ref{eqn:c_and_q}) to be zero whenever 
$\widetilde{\chi}(k) = 0$, leading to the requirement
\begin{equation}
\widetilde{Q}(k) = \frac{1}{\rho \widetilde{m}^2(k)} . \label{eqn:q_value}
\end{equation}
Inserting Eq.~(\ref{eqn:q_value}) into Eq.~(\ref{eqn:fourier_oz_equation}) 
gives the one-component decorated OZ equation:
\begin{eqnarray}
\widetilde{H}(k) & = & \widetilde{C}(k) + \frac{\widetilde{C}(k) \widetilde{H}(k)}{\rho \widetilde{m}^2(k)} , \\
\widetilde{C}(k) & = & \rho \widetilde{m}^2(k) - \frac{\rho^2 \widetilde{m}^4(k)}{\widetilde{\chi}(k)} . \label{eqn:one_component_OZ}
\end{eqnarray}

Relation (\ref{eqn:one_component_OZ}) holds for a decorated single-component
system.
The generalization of Eq.~(\ref{eqn:one_component_OZ}) for a 
multiple-component system can be obtained by noting that
$\widetilde{Q}^{-1}(k)$ is equal to the self-correlation term.
For example, for a two-component system of nonoverlapping spheres,
the relations analogous to
(\ref{eqn:q_value})--(\ref{eqn:one_component_OZ}) are given by
\begin{eqnarray}
\widetilde{Q}(k) & = & \frac{1}{\rho_A \widetilde{m}_A^2(k) + \rho_B \widetilde{m}_B^2(k)} , \\
\widetilde{H}(k) & = & \widetilde{\chi}(k) - \rho_A \widetilde{m}_A^2(k) - \rho_B \widetilde{m}_B^2(k) = \widetilde{C}(k) + \frac{\widetilde{C}(k) \widetilde{H}(k)}{\rho_A \widetilde{m}_A^2(k) + \rho_B \widetilde{m}_B^2(k)} , \\
\widetilde{C}(k) & = & \rho_A \widetilde{m}_A^2(k) + \rho_B \widetilde{m}_B^2(k) - \frac{\left(\rho_A \widetilde{m}_A^2(k) + \rho_B \widetilde{m}_B^2(k)\right)^2}{\widetilde{\chi}(k)} , \label{eqn:c_two_components}
\end{eqnarray}
where $\rho_A$ and $\rho_B$ are the number densities of species $A$ and $B$,
respectively, and $m_A(r)$ and $m_B(r)$ are the corresponding sphere indicator
functions.

\subsubsection{Mixture Case \label{sec:mixture_OZ}}

Consider an $M$-component system, in which $N_\alpha$ represents the number
of particles of species $\alpha$, where $\alpha = A, B, \ldots$
Following Ref. \onlinecite{Baxter1970}, we write the following OZ equation 
for the mixture total correlation function $h_{\alpha\beta}(r)$ and the
direct correlation function $c_{\alpha\beta}(r)$:
\begin{equation}
h_{\alpha\beta}(r) = c_{\alpha\beta}(r) + \sum_{\gamma=1}^M \rho_\gamma c_{\alpha\gamma} \otimes h_{\gamma\beta}(r) , \label{eqn:mixture_raw_OZ}
\end{equation}
where $\alpha$, $\beta$, and $\gamma$ represent the different components of
the system.
Note that $c_{\alpha\beta}(r)$ is different from
the ``decorated'' ``two-phase'' direct correlation function $C(r)$ defined in
Sec.~\ref{sec:single_component_case}.
Equation (\ref{eqn:mixture_raw_OZ}) can be rewritten in matrix form:
\begin{eqnarray}
\sqrt{\rho_\alpha \rho_\beta} h_{\alpha\beta}(r) & = & \sqrt{\rho_\alpha \rho_\beta} c_{\alpha\beta}(r) + \sum_\gamma \sqrt{\rho_\alpha \rho_\gamma} c_{\alpha\gamma}(r) \otimes \sqrt{\rho_\gamma \rho_\beta} h_{\gamma\beta}(r) \nonumber \\
\mathbf{H}(r) & = & \mathbf{C}(r) + \mathbf{C}(r) \otimes \mathbf{H}(r), \label{eqn:matrix_raw_OZ}
\end{eqnarray}
where the components of the matrices $\mathbf{H}(r)$ and $\mathbf{C}(r)$ are 
given by
\begin{eqnarray}
H_{\alpha\beta}(r) & = & \sqrt{\rho_\alpha \rho_\beta} h_{\alpha\beta}(\mathbf{r}) , \\
C_{\alpha\beta}(r) & = & \sqrt{\rho_\alpha \rho_\beta} c_{\alpha\beta}(\mathbf{r}) .
\end{eqnarray}
Taking the Fourier transform of Eq.~(\ref{eqn:matrix_raw_OZ}) gives
\begin{eqnarray}
\widetilde{\mathbf{H}}(k) & = & \widetilde{\mathbf{C}}(k) + \widetilde{\mathbf{C}}(k) \widetilde{\mathbf{H}}(k) , \\
\widetilde{\mathbf{C}}(k) & = & \widetilde{\mathbf{H}}(k) \left( \mathbf{I} + \widetilde{\mathbf{H}}(k) \right)^{-1} , \label{eqn:bigC_with_H}
\end{eqnarray}
where $\mathbf{I}$ is the identity matrix.

Equation (\ref{eqn:bigC_with_H}) can be simplified by introducing the
$M \times M$
multiple-component structure factor matrix $\mathbf{S}(k)$, whose components
are denoted as $S_{\alpha\beta}(k)$:
\begin{eqnarray}
\mathbf{S}(k) & = &
\left( \begin{array}{ccc}
S_{AA}(k) & S_{AB}(k) & \cdots \\
S_{AB}^*(k) & S_{BB}(k) & \cdots \\
\vdots & \vdots & \ddots
\end{array} \right)
 = 
\left( \begin{array}{ccc}
1 + \rho_A \widetilde{h}_{AA}(k) & \sqrt{\rho_A \rho_B} \widetilde{h}_{AB}(k) & \cdots \\
\sqrt{\rho_A \rho_B} \widetilde{h}_{AB}^*(k) & 1 + \rho_B \widetilde{h}_{BB}(k) & \cdots \\
\vdots & \vdots & \ddots
\end{array} \right) , \nonumber \\
 & = & \mathbf{I} + \widetilde{\mathbf{H}}(k) \label{eqn:S_from_H},
\end{eqnarray}
where $S^*_{\alpha\beta}(k)$ denotes the complex conjugate of
$S_{\alpha\beta}(k)$.
Substitution of Eq.~(\ref{eqn:S_from_H}) into Eq.~(\ref{eqn:bigC_with_H})
yields the following simpler expression for $\widetilde{\mathbf{C}}(k)$:
\begin{equation}
\widetilde{\mathbf{C}}(k) = \mathbf{I} - \mathbf{S}(k)^{-1} .
\end{equation}
This last equation should be used carefully, since the $\mathbf{S}(k)$ matrix
is rank-1 for a single realization of a system, and hence it cannot be
inverted without first taking an \emph{ensemble average}.\cite{footnote:rank-1}

Equation (\ref{eqn:c_two_components}), valid for the ``two-phase'' decoration, 
and Eq.~(\ref{eqn:bigC_with_H}) may not look
similar, but their similarities can be made apparent by rewriting
$\widetilde{\chi}(\mathbf{k})$ and $S_{\alpha\beta}(\mathbf{k})$ in terms of the
collective coordinates $\widetilde{\rho}_\alpha(\mathbf{k})$:
\begin{equation}
\widetilde{\rho}_\alpha(\mathbf{k}) = \sum_{j=1}^{N_\alpha} e^{i \mathbf{k} \cdot \mathbf{r}_j^\alpha},
\end{equation}
where $\mathbf{k}$ is the wave vector and $N_\alpha$ is the number of particles
of species $\alpha$ 
For a single configuration of a multiple-component system in a volume $V$, 
we get the structure factor matrix components to be given by
\begin{equation}
S_{\alpha\beta}(\mathbf{k}) = \frac{\widetilde{\rho}_\alpha(\mathbf{k}) \widetilde{\rho}^*_\beta(\mathbf{k})}{\sqrt{N_\alpha N_\beta}} - V \delta_{\mathbf{k}, \mathbf{0}}. \label{eqn:S_using_coll_coords}
\end{equation}
Since we never compute $S_{\alpha\beta}(k = 0)$ directly, instead
relying on the $k \rightarrow 0$ limit, we can drop the
Kronecker delta function in the following steps.
For a two-component system, the spectral density for the decorated system is
\begin{eqnarray}
\chi(\mathbf{r}) & = & \rho_A m_A(\mathbf{r}) \otimes m_A(\mathbf{r}) + \rho_B m_B(\mathbf{r}) \otimes m_B(\mathbf{r}) + \\
 & & \rho_A^2 m_A(\mathbf{r}) \otimes h_{AA}(\mathbf{r}) \otimes m_A(\mathbf{r}) + \rho_A \rho_B m_A(\mathbf{r}) \otimes h_{AB}(\mathbf{r}) \otimes m_B(\mathbf{r}) + \nonumber \\
 & & \rho_A \rho_B m_B(\mathbf{r}) \otimes h_{BA}(\mathbf{r}) \otimes m_A(\mathbf{r}) + \rho_B^2 m_B(\mathbf{r}) \otimes h_{BB}(\mathbf{r}) \otimes m_B(\mathbf{r}) \nonumber ,
\end{eqnarray}
for which the Fourier transform is given by
\begin{eqnarray}
\widetilde{\chi}(\mathbf{k}) & = & \rho_A \widetilde{m}_A^2(\mathbf{k})  + \rho_B \widetilde{m}_B^2(\mathbf{k})  + \rho_A^2 \widetilde{m}_A^2(\mathbf{k}) \widetilde{h}_{AA}(\mathbf{k})  + \rho_A \rho_B \widetilde{m}_A(\mathbf{k})  \widetilde{m}_B(\mathbf{k})  \widetilde{h}_{AB}(\mathbf{k})  + \nonumber \\
& & \rho_A \rho_B \widetilde{m}_A(\mathbf{k}) \widetilde{m}_B(\mathbf{k}) \widetilde{h}_{BA}(\mathbf{k}) + \rho_B^2 \widetilde{m}_B^2(\mathbf{k}) \widetilde{h}_{BB}(\mathbf{k}) \nonumber \\
 & = & \frac{\left| \widetilde{\rho}_A(\mathbf{k}) \widetilde{m}_A(\mathbf{k}) + \widetilde{\rho}_B(\mathbf{k}) \widetilde{m}_B(\mathbf{k}) \right|^2}{V} . \label{eqn:chi_using_rho}
\end{eqnarray}
Using Eq.~(\ref{eqn:chi_using_rho}) to rewrite Eq.~(\ref{eqn:c_two_components})
leads to
\begin{equation}
\widetilde{C}(\mathbf{k}) = \rho_A \widetilde{m}_A^2(\mathbf{k}) + \rho_B \widetilde{m}_B^2(\mathbf{k}) \left(1 - \frac{N_A \widetilde{m}_A^2(\mathbf{k}) + N_B \widetilde{m}_B^2(\mathbf{k})}{\left| \widetilde{\rho}_A(\mathbf{k}) \widetilde{m}_A(\mathbf{k}) + \widetilde{\rho}_B(\mathbf{k}) \widetilde{m}_B(\mathbf{k}) \right|^2} \right) . \label{eqn:c_two_component_rho}
\end{equation}

Now, assume that the decoration of the two-component system is chosen 
such that $\psi(\mathbf{k}) = \left( \sqrt{\rho_A} \widetilde{m}_A(\mathbf{k}), 
\sqrt{\rho_B} \widetilde{m}_B(\mathbf{k}) \right)^\top$ is an eigenvector of 
$\mathbf{S}(\mathbf{k})$. Calculating the associated eigenvalue of $\widetilde{\mathbf{C}}(\mathbf{k})$
(which shares eigenvectors with $\mathbf{S}(\mathbf{k})$) leads to
\begin{equation}
\frac{\psi^{*\top}(\mathbf{k}) \widetilde{\mathbf{C}}(\mathbf{k}) \psi(\mathbf{k})}{\rho_A \widetilde{m}_A^2(\mathbf{k}) + \rho_B \widetilde{m}_B^2(\mathbf{k}) } =
1 - \frac{N_A \widetilde{m}_A^2(\mathbf{k}) + N_B \widetilde{m}_B^2(\mathbf{k})}{\left< \left| \widetilde{\rho}_A(\mathbf{k}) \widetilde{m}_A(\mathbf{k}) + \widetilde{\rho}_B(\mathbf{k}) \widetilde{m}_B(\mathbf{k}) \right|^2 \right>} , \label{eqn:c_matrix_eigenvalue}
\end{equation}
The similarities between Eqs.~(\ref{eqn:c_two_component_rho}) and
(\ref{eqn:c_matrix_eigenvalue}) are striking, and lend credibility to their
use. However, it should not be forgotten that 
Eq.~(\ref{eqn:c_matrix_eigenvalue}) is only valid for a very precise choice of
$\widetilde{m}_A(\mathbf{k})$ and $\widetilde{m}_B(\mathbf{k})$, which may or may not
be realizable for arbitrary systems. It is therefore more appropriate to
use a decoration that uses \emph{a priori} information 
about the system (e.g. an effective radius of the particles) together with 
Eq.~(\ref{eqn:c_two_component_rho}). In a situation where such
information is missing, calculating the actual eigenvalues of 
$\mathbf{S}(\mathbf{k})$ and $\widetilde{\mathbf{C}}(\mathbf{k})$ is a good 
alternative choice, although it requires multiple realizations of the
system in order to get the ensemble-average values.

\section{Simulation Details \label{sec:sim_details}}

We carry out molecular dynamics simulations in the $NVT$ ensemble to
study the behavior of two different atomic glass-forming liquid models:
a three-dimensional single-component system in which the particles
interact with the Z2 Dzugutov potential and a two-dimensional
two-component system in which the particles interact with the Kob-Andersen
potential. In particular, starting from liquid states, 
we quench these two model systems and follow their transitions
from fluids, to supercooled fluids and glassy states and determine their
associated supercooled and glassy states as a function of temperature.

The interacting systems consist of $N = 100000$ particles in
a two-dimensional (Kob-Andersen) or three-dimensional (Z2 Dzugutov) 
periodic box, subject to a Nos\'e-Hoover thermostat\cite{Thermostat_reference}
with a mass set to $N / 1000 = 100$. This particular choice of mass is
selected to avoid the numerical instabilities that occur when a small
mass is used, while reducing the time the thermostat takes to equilibrate
which increases with larger masses.
The initial configurations are generated using
the random sequential addition (RSA) algorithm,\cite{RSA_reference} and with
an initial temperature that is much larger than the freezing temperature.
There are four relevant units in the molecular dynamics simulations:
units of energy, length, mass, and time, of which three can chosen
independently. The units of energy and length are selected by the numerical
values of the potentials' parameters, while the unit of mass is set by
letting all particles have unit masses. These choices defined the
natural units, including the unit of time.
The system is then continuously cooled using an exponential rate
\begin{equation}
T(t) = T_0 \times 10^{-t / \tau_{10}},
\end{equation}
where $T(t)$ is the temperature when the simulation has been running for a
time $t$, $T_0$ is the initial temperature, and
the time per decade $\tau_{10}$ controls the cooling rate.
The molecular dynamics integration is done using the velocity Verlet scheme.

For the Z2 Dzugutov potential, shown in Eq.~(\ref{DZ}), we use the following parameter  values: $a = 1.04$, $\eta = 0.33$, $k_F = 4.139$, $b = 4.2 \times 10^7$, $\sigma = 0.348$, $n = 14.5$, $r_c = 2.64488$, and $V_0 = 0.13391543$. The values of $r_c$ and $V_0$ are chosen such that both $v(r_c) = 0$ and $\left. \frac{dv}{dr} \right|_{r=r_c} = 0$.
This choice of parameters defines the natural units of both energy and length.
Following Ref.~\onlinecite{Do03}, the particle density is fixed as 
$\rho = 0.84$ and the particle mass is set to unity.
The time per decade $\tau_{10}$ is set to $500$, $200$, and $50$ natural time 
units. Slower cooling schedules are attempted (such as $\tau_{10} = 2000$),
but they lead to some of the samples crystallizing.
The time step is $\Delta t = 5 \times 10^{-3}$ in the natural time units
and is chosen such that the total energy of the system is conserved when
the thermostat is removed.

For the Kob-Andersen potential, shown in Eq.~(\ref{KA}), 
we use a composition of particles with number ratio $A:B=65:35$ and the
following parameters: $\sigma_{AA}=1.0$, $\epsilon_{AA}=1.0$, 
$\sigma_{AB}=\sigma_{BA}=0.8$, $\epsilon_{AB} = \epsilon_{BA} = 1.5$,
$\sigma_{BB} = 0.88$, and $\epsilon_{BB} = 0.5$. The values for the
$V_{0 \alpha \beta}$ are chosen such that the potentials are continuous at 
$r = 2.5 \sigma_{\alpha \beta}$ cutoffs. 
These choices of parameters define the natural units of energy
($\epsilon_{AA}$) and length ($\sigma_{AA}$).
Both particle species are assumed to have
masses equal to unity. Following Ref. \onlinecite{BBBKMR2007a}, 
we set $\rho = 1.161662$. The time per decade of temperature decay $\tau_{10}$
is set to $2000$, $400$, $100$, and $20$.
The time step is $\Delta t = 1 \times 10^{-3}$.

\section{Results \label{sec:results}}

\subsection{Z2 Dzugutov Single-Component Glass}

To estimate the glass transition temperature $T_g$ of the Z2 Dzugutov
model, we use the temperature at which the total energy per particle as
a function of temperature changes slope most rapidly.
Since the harmonic contribution $3 k_B T$ to the average total energy
per particle $u$ has a
constant slope, we subtract it from $u$ to detect any change of slope.
As seen in Fig.~\ref{fig:Z2_glass_transition},
we obtain $k_B T_g \sim 0.88$ for the Z2 Dzugutov model.
Comparatively, by observing the highest temperature at which the supercooled
systems crystallized and the temperature at which such crystals melt,
we roughly estimate the melting temperature to be $T_m / T_g \sim 2.5 \pm 0.5$.

\begin{figure}[htp]
	\centering
	\includegraphics[scale=0.3]{Fig1.eps}
	\caption{(Color online) Strictly anharmonic portion of the total 
	average energy
	(kinetic and potential) per particle $u - 3k_B T$
	of the system in term of the thermostat temperature $T$.
	This is obtained by averaging over 10 cooling simulations of
	supercooled Z2 Dzugutov systems using $\tau_{10} = 400$.
	$3 k_B T$ has been subtracted from the energy to help identify the glass
	transition.
	The glass transition temperature $k_B T_g \sim 0.88$ is estimated by 
	finding the	temperature at which the function slope changes most rapidly.
	The vertical dashed line is located at $T = T_g$.
	The energy scale is normalized through our choice of potential parameters
	(see Sec.~\ref{sec:sim_details}).
	}
	\label{fig:Z2_glass_transition}
\end{figure}

\begin{figure}[htp]
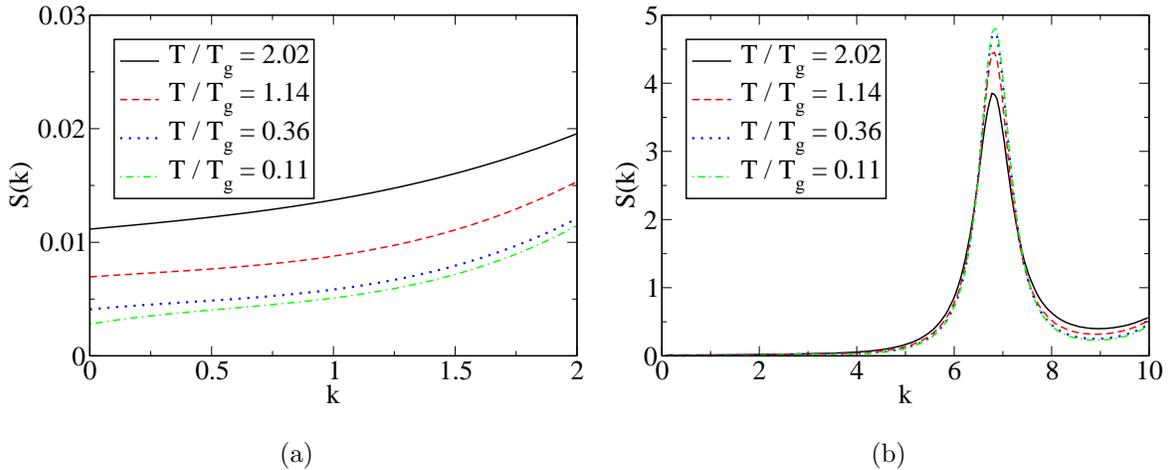

	\centering
	\subfigure[]{
		\includegraphics[scale=0.3]{Fig2a.eps}
		\label{fig:Z2_Sk_small_k}
	}
	\subfigure[]{
		\includegraphics[scale=0.3]{Fig2b.eps}
		\label{fig:Z2_Sk_large_k}
	}
	\caption{(Color online) Structure factors $S(k)$ for Z2 Dzugutov systems
	supercooled using $\tau_{10} = 500$ for various temperatures. 
	The curves have been averaged over 10 realizations.
	\subref{fig:Z2_Sk_small_k} Cubic fits of the small-wavenumber
	($k < 2$) structure factors. The type of fits and their cutoff are
	chosen such that they accurately reproduce the
	features of the structure factors, in particular the positive linear
	dependence near $k = 0$.
	\subref{fig:Z2_Sk_large_k} Larger-wavenumber structure factors.
	}
	\label{fig:Z2_Sk}
\end{figure}

To calculate the volume integral of the direct correlation function $c(r)$,
we need to find the limit of $S(k)$ for $k \rightarrow 0$, and then
substitute it in Eq.~(\ref{Ck}).
Since $S(k=0)$ cannot be calculated directly
in a finite simulation box of side length $L$
because the smallest possible wavenumber accessible is $2\pi/L$,
an extrapolation from the available data to zero wavenumber must be used.
Figure \ref{fig:Z2_Sk_small_k} shows the small-wavelength
behavior of $S(k)$ for Z2 Dzugutov model at different temperatures.
It is clear that $S(k)$ is nearly linear in $k$ for $k \lesssim 1$, 
leading to a very good fit to a linear function. This linear behavior of
$S(k)$ for small $k > 0$ implies that the real-space total correlation
function $h(r)$ decays, for large but finite $r$, as a power law
$-1 / r^4$ or, equivalently, the direct correlation function decays
as $c(r) \sim -1/r^2$.
The numerical value of $S(k=0)$ only changes by up to 5\% between the cubic fit
for $k < 2$ shown in Fig.~\ref{fig:Z2_Sk_small_k} and a linear fit for $k < 1$.
Since the linear fit is less susceptible to overfitting and complex behavior
for $1 < k < 2$, we elect to use this linear fit to
extrapolate the value of $S(k=0)$ for these systems.

\begin{figure}[htp]
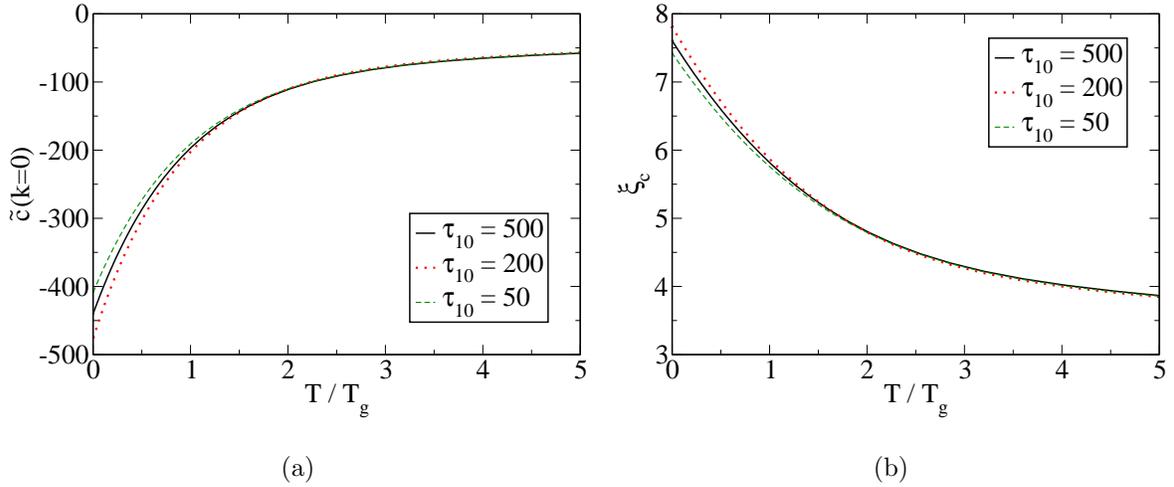

	\centering
	\subfigure[]{ 
		\includegraphics[scale=0.3]{Fig3a.eps}
		\label{fig:Z2_ck_sub}
	}
	\subfigure[]{
		\includegraphics[scale=0.3]{Fig3b.eps}
		\label{fig:Z2_lengthscale}
	}
	\caption{(Color online) Growing length scales for Z2 Dzugutov systems
	generated using various cooling schedules. For each cooling schedule, the
	results have been averaged over 10 realizations and fitted to
	the sum of an exponential and a linear function to smooth out the
	numerical noise.
	\subref{fig:Z2_ck_sub} Limit of $\widetilde{c}(k)$ for $k \rightarrow 0$,
	calculated using linear fits of $S(k)$.
	\subref{fig:Z2_lengthscale} The static length scale $\xi_c$,
	defined by relation (\ref{eqn:xi_Z2}), associated with these systems.
	Note that the nearest neighbor distance between particles at 
	$T=0$ is $1.0539$.
	}
	\label{fig:Z2_ck}
\end{figure}

From the Fourier transform of the direct correlation function $\widetilde{c}(k)$,
which has units of volume, we define the following length scale:
\begin{equation}
\xi_c \equiv \left[ -\widetilde{c}(0) \right]^{1/d} , \label{eqn:xi_Z2}
\end{equation}
where $d$ is the Euclidean dimension.
From Fig.~\ref{fig:Z2_ck_sub}, there is a striking evidence that
$\widetilde{c}(k=0)$ grows to a large negative value in the supercooled regime,
leading to a doubling in the value of the length scale $\xi_c$.

\begin{figure}[htp]
	\centering
	\includegraphics[scale=0.3]{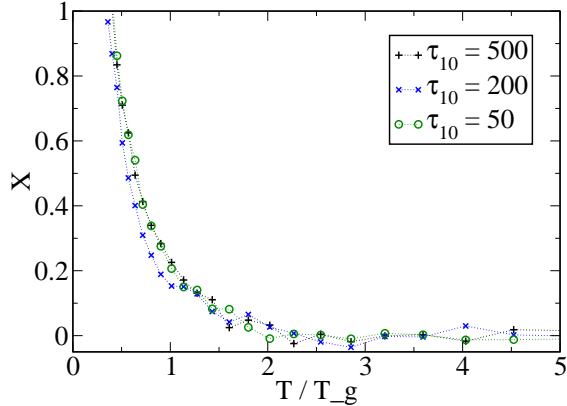}
	\caption{(Color online)
	Nonequilibrium index $X$ for Z2 Dzugutov systems supercooled
	using various cooling schedules defined in Eq.~(\ref{eqn:X_def}).
	}
	\label{fig:Z2_X}
\end{figure}

In the case of a single-component system at equilibrium,
the compressibility relation links
its isothermal compressibility $\kappa_T = -\frac{1}{V} \left.\frac{\partial V}{\partial p}\right|_T$
to its structure factor as follows:
\begin{equation}
\rho \kappa_T k_B T = S(0). \label{eqn:comp_relation}
\end{equation}
However, supercooled liquids and glasses are not equilibrium states and
consequently Eq.~(\ref{eqn:comp_relation}) tends not to be satisfied. 
Following Ref.~\onlinecite{Ho12}, we use the
deviation from Eq.~(\ref{eqn:comp_relation}) to measure a
nonequilibrium index $X$:
\begin{equation}
X \equiv \frac{S(0)}{\rho \kappa_T k_B T} - 1. \label{eqn:X_def}
\end{equation}
The isothermal compressibility $\kappa_T$ is computed by the following
finite difference formula:
\begin{equation}
\kappa_T \simeq - \frac{\Delta V}{V} \frac{1}{\Delta P}, \label{eqn:kappa_T}
\end{equation}
where $\Delta V$ is the change in volume of the simulation box and
$\Delta P$ is the resulting change in pressure of the system after it is
allowed to relax at constant temperature. The pressure is calculated using
the virial relation. It bears mentioning that since the system is not at
equilibrium, it is not in a steady state even before the change in volume.
To minimize the impact of the uncompressed system relaxation, both
the uncompressed and compressed systems are allowed to relax for the
same amount of time before measuring their pressures.

In the case of the Z2 Dzugutov system, we use a change of volume
$\Delta V / V = 0.3\%$ and the pressure is sampled from $t = 5$ to
$t = 10$, where $t = 0$ denotes the time at which the system is compressed.
As can be seen in Fig.~\ref{fig:Z2_X}, $X$ is 
zero\cite{footnote:zeroX} for $T / T_g > 2$,
with only slight deviations due to noise and numerical inaccuracies.
However, as the temperature is lowered to values approaching the glass 
transition, $X$ increases up to a value of $\sim 0.2$ at $T / T_g = 1$.
For $T < T_g$, the inability of the system to relax in a time of the
order of the cooling schedule time per decade $\tau_{10}$
results in nearly constant values of $\kappa_T$ and $S(0)$
which leads to the asymptotic behavior of $X$ as $T \rightarrow 0$.

\begin{figure}[htp]
	\centering
	\includegraphics[scale=0.3]{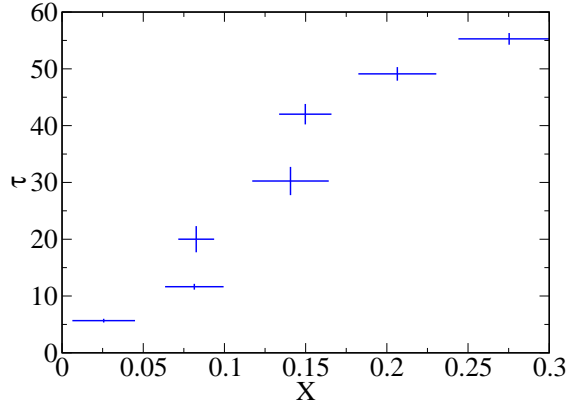}
	\caption{(Color online)
	Timescale $\tau$ of the early relaxation process of the system
	versus the nonequilibrium index $X$. Both quantities have been averaged
	over 10 configurations.
	The circle are centered on the averages of $X$ and $\tau$, while the
	horizontal and vertical lines represent their respective uncertainties,
	with their half-length set equal to the average standard deviations.
	The initial configurations which are
	allowed to relax at constant temperature are generated from the liquid
	phase through a cooling schedule employing $\tau_{10} = 50$.
	Each datum represents a single temperature.
	Observe that $\tau$ and $X$ are positively correlated.
	Therefore, since $X$ is a monotonically decreasing function of the 
	temperature $T$ (see Fig~\ref{fig:Z2_X}), $\tau$ also increases with
	decreasing $T$.
	The values of $T / T_g$ associated with each datum are, in order of 
	smallest to largest $\tau$ are as follows:
	1.80, 1.61, 1.43, 1.28, 1.14, 1.01, and 0.90.
	}
	\label{fig:Z2_relaxation}
\end{figure}

Is the growing nonequilibrium index $X$, a purely static quantity, correlated
with the growing relaxation times as the temperature decreases during the
supercooling process? Figure \ref{fig:Z2_relaxation} shows a positive
correlation between $X$ and $\tau$, where $\tau$ is the timescale associated 
with the early relaxation process, extracted from an exponential fit function
$\sim e^{-t / \tau}$ of the system total energy.
To observe this process, we start with configurations that have been
supercooled to a given temperature following a specific cooling schedule.
These configurations are then allowed to evolve at constant temperature.
It can be clearly seen that $X$ and $\tau$ are strongly
and positively correlated.

\subsection{Kob-Andersen $A_{65}B_{35}$ Two-Component Glass}

\begin{figure}[htp]
	\centering
	\includegraphics[scale=0.6]{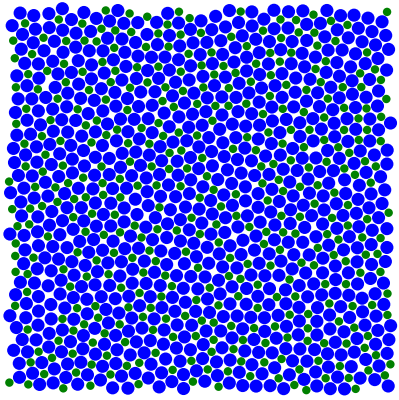}
	\caption{(Color online) Example of a decorated Kob-Andersen glass
	configuration (a small subregion of the configuration only). The larger
	disks represent the $A$ particles, while the smaller disks represent B 
	particles. The radii of the disks are chosen such that the two closest
	$A$ particles of the whole configuration are in contact and the closest
	$A$--$B$ pair of particles are in contact. The configuration shown has
	been generated using $\tau_{10} = 100$, and is at a temperature of
	$T/T_g= 6.7 \times 10^{-5}$. The particle radii are $R_A=0.513720$ and
	$R_B=0.329883$ ($R_A/R_B = 1.55728$).
	}
	\label{fig:KA_configuration}
\end{figure}

To calculate the spectral density $\widetilde{\chi}(k)$, we decorate the 
systems by circumscribing disks of radius $R_A$ and $R_B$ centered around the 
point particles of species $A$ and $B$, respectively. Since our derivation in 
Sec.~\ref{sec:two_phase_decoration} requires the disks to be 
nonoverlapping,
we chose the largest possible radii that satisfy this condition. In the case
of a Kob-Andersen glass, $A$ particles are often located next to one another,
while $B$ particles can be further apart. 
This leads to our decision to use the
distances between the closest $A$--$A$ and $A$--$B$ pairs of particles to 
define the particle radii. Figure \ref{fig:KA_configuration} shows part
of a glass configuration decorated using this procedure.

\begin{figure}[htp]
	\centering
	\includegraphics[scale=0.3]{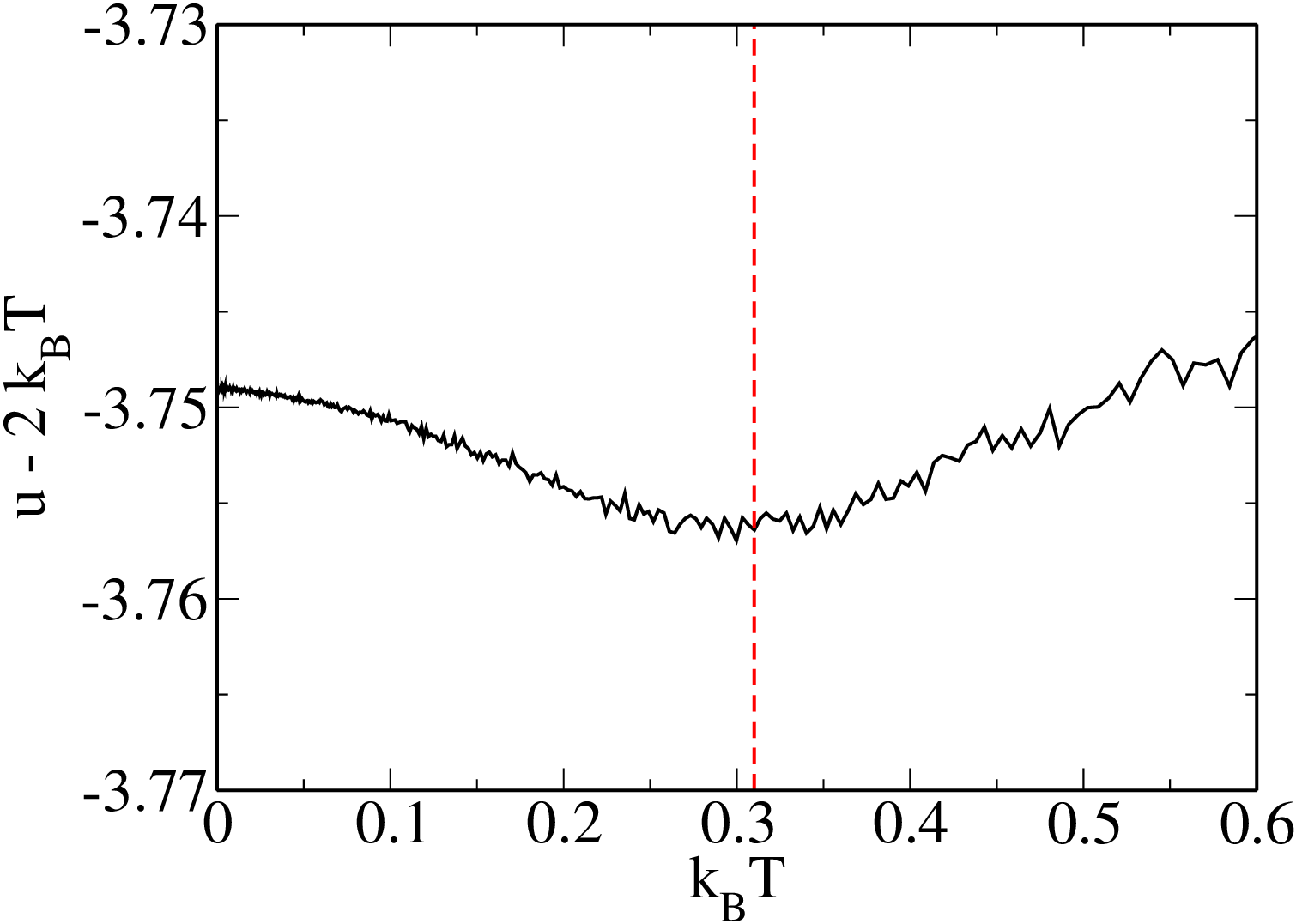}
	\caption{(Color online) Strictly anharmonic portion of the total 
	average energy (kinetic and potential) per particle $u - 2k_B T$
	of the system in terms of the thermostat temperature $T$.
	This is obtained by averaging 10 cooling simulations of
	supercooled Kob-Andersen systems using $\tau_{10} = 400$.
	$2 k_B T$ has been subtracted from the energy to help identify
	the glass transition.
	The glass transition temperature $T_g \sim 0.31$ is estimated by 
	finding the	temperature at which the function slope changes the most
	rapidly. The vertical dashed line is located at $T = T_g$.
	The energy scale is normalized through our choice of potential parameters
	(see Sec.~\ref{sec:sim_details}).
	}
	\label{fig:KA_glass_transition}
\end{figure}

In an identical fashion to the Z2 Dzugutov system, we use the change in slope
of the total energy in terms of the temperature to estimate the glass
transition temperature $T_g$ for the Kob-Andersen system.
Since the Kob-Andersen system that we analyze is two-dimensional, its harmonic
contribution to the energy is $2 k_B T$, which we subtract from the 
total average energy per particle $u$ to detect any change of slope.
The result obtained from Fig.~\ref{fig:KA_glass_transition} is $T_g \sim 0.31$,
which is reasonably close to the previously-reported value of $T_g = 0.33$.
\cite{Bruning2009}

\begin{figure}[htp]
	\centering
	\includegraphics[scale=0.3]{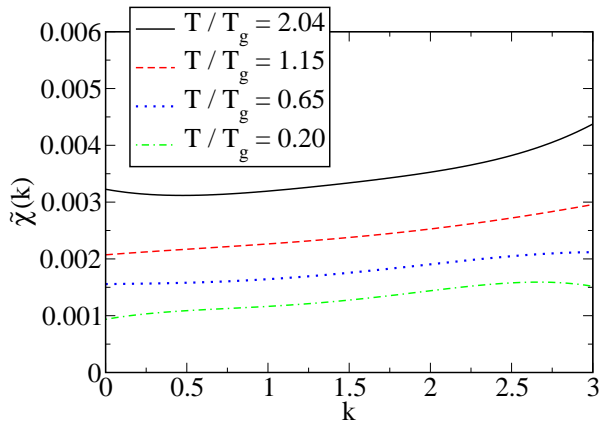}
	\caption{(Color online) Spectral density $\widetilde{\chi}(k)$ versus
	wavenumber $k$ for
	Kob-Andersen $A_{65}B_{35}$ systems supercooled using $\tau_{10}=400$.
	The curves have been averaged over 10 realizations and fitted using
	fourth degree polynomials. The type of fits have been chosen for their
	ability to reproduce accurately the features of the structure factors for
	the range presented ($0 < k < 3$).
	The disk radii for the decorations are calculated independently for each
	configuration.
	}
	\label{fig:KA_chik}
\end{figure}

\begin{figure}[htp]
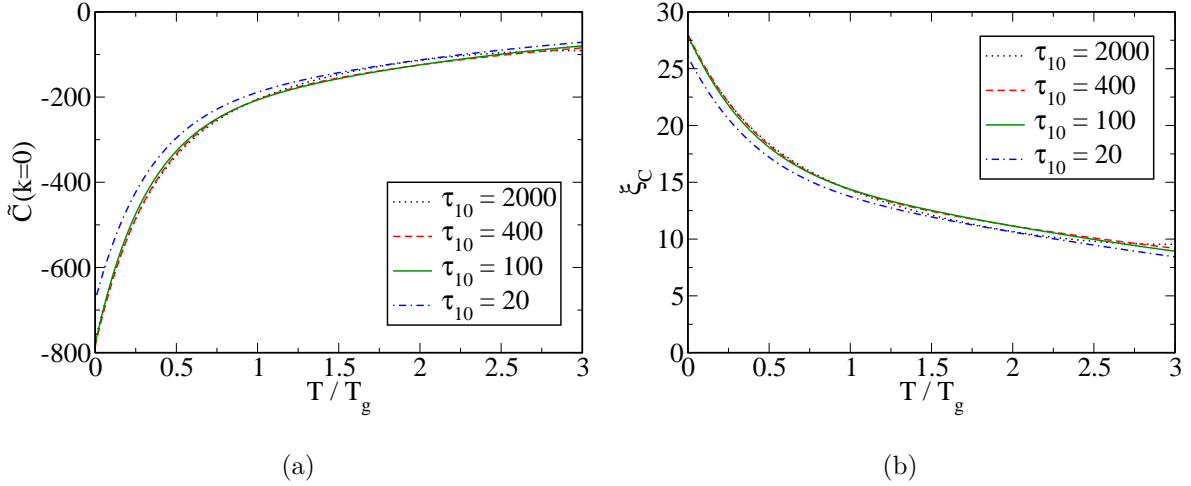

	\centering
	\subfigure[]{
		\includegraphics[scale=0.3]{Fig9a.eps}
		\label{fig:KA_ck_sub}
	}
	\subfigure[]{
		\includegraphics[scale=0.3]{Fig9b.eps}
		\label{fig:KA_lengthscale}
	}
	\caption{(Color online) Growing length scales for two-dimensional 
	Kob-Andersen systems.
	For each cooling schedule, the results have been averaged over 10 
	realizations and fitted to the sum of an exponential and a
	quadratic functions to smooth out the numerical noise.
	\subref{fig:KA_ck_sub} Limit of $\widetilde{C}(k)$ for $k \rightarrow 0$,
	calculated using the linear fits of $\widetilde{\chi}(k)$.
	\subref{fig:KA_lengthscale} The static length scale $\xi_C$,
	defined by relation (\ref{eqn:xi_KA}), associated with these systems.
	}
	\label{fig:KA_ck}
\end{figure}

As in the case of Z2 Dzugutov systems, the spectral densities
$\widetilde{\chi}(k)$ for Kob-Andersen liquids, supercooled liquids, and glasses
have nearly linear behavior for
$k \lesssim 1$. It is thus possible to prescribe a linear fit to
extrapolate the values of $\widetilde{\chi}(k=0)$, which is required to calculate
$\widetilde{C}(k=0)$ using Eq.~(\ref{eqn:c_two_components}).
We again define a length scale based on the $\widetilde{C}(k=0)$:
\begin{equation}
\xi_C \equiv \left[ -\widetilde{C}(0) \right]^{1/d} , \label{eqn:xi_KA}
\end{equation}
where $d$ is the Euclidean dimension.
Figure \ref{fig:KA_ck_sub} shows the large change in value of
$\widetilde{C}(k=0)$ as the Kob-Andersen liquids are supercooled,
leading to the length scale $\xi_C$ to increase
by a factor larger than 5 between the fluid states and the zero-temperature
glassy states.

\begin{figure}[htp]
	\centering
	\includegraphics[scale=0.3]{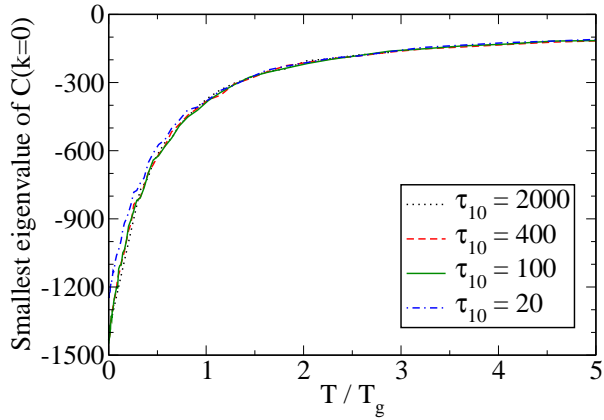}
	\caption{(Color online) Smallest eigenvalue
	of $\lim_{k \rightarrow 0} \mathbf{C}(k)$, 
	calculated using a linear fit of the
	matrix structure factor $\mathbf{S}(k)$. While the qualitative behavior of
	this eigenvalue can be compared to $\widetilde{C}(k=0)$ 
	(see Fig.~\ref{fig:KA_ck_sub}), their quantitative values cannot directly
	be compared because they have different units: 
	$\widetilde{C}(k)$ has units of volume, while $\widetilde{\mathbf{C}}(k)$ is 
	dimensionless.
	}
	\label{fig:KA_eigen}
\end{figure}

As mentioned in Sec.~\ref{sec:mixture_OZ}, there is a second generalization
of the direct correlation function which does not require any \emph{a priori}
knowledge or about the particle shapes. Instead, one can use the
matrix direct correlation function $\mathbf{C}(r)$ and its Fourier transform
$\widetilde{\mathbf{C}}(k)$. As can be observed in Fig.~\ref{fig:KA_eigen}, the qualitative
behavior of the smallest eigenvalue of $\widetilde{\mathbf{C}}(k)$ in the $k \rightarrow 0$
limit is strikingly close to the behavior of $\widetilde{C}(k)$ in the
same limit. This indicates that our decoration choice is
appropriate for detecting long-range density fluctuations in Kob-Andersen
glasses and supercooled liquids.

\begin{figure}[htp]
	\centering
	\includegraphics[scale=0.3]{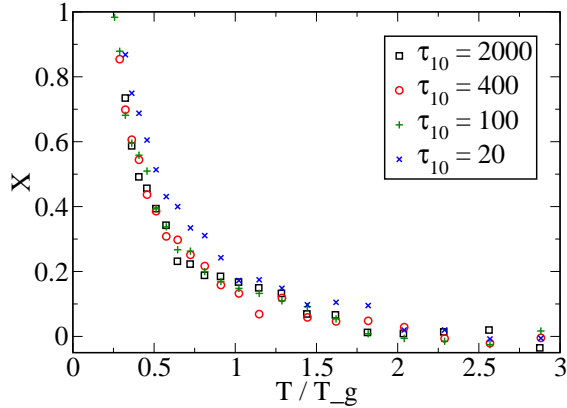}
	\caption{(Color online)
	Nonequilibrium index $X$ for Kob-Andersen systems supercooled using
	various cooling schedules defined in Eq.~(\ref{eqn:X_def_KA}).
	}
	\label{fig:KA_X}
\end{figure}

Since the compressibility relation (\ref{eqn:comp_relation}) applies only to
single-component systems, we must generalize the nonequilibrium index $X$ for
mixtures. The compressibility relation for multicomponent systems at
equilibrium, is given by:\cite{Kirkwood1951}
\begin{equation}
\kappa_T k_B T = \frac{|\mathbf{B}|}{\sum_{\alpha=1}^M \sum_{\beta=1}^M |\mathbf{B}|_{\alpha\beta}} , \label{eqn:kirkwood_compress}
\end{equation}
where the components $B_{\alpha\beta}$ of the matrix $\mathbf{B}$ are
\begin{equation}
B_{\alpha\beta} = \frac{\sqrt{N_\alpha N_\beta}}{V} \lim_{k \rightarrow 0} S_{\alpha\beta}(k),
\end{equation}
$|\mathbf{B}|$ is the determinant of $\mathbf{B}$, and 
$|\mathbf{B}|_{\alpha\beta}$ is the $\alpha\beta$ minor of $\mathbf{B}$.
The nonequilibrium index $X$ for multicomponent systems can now be
defined by using the mismatch between the left and right sides of
Eq.~(\ref{eqn:kirkwood_compress}), that is,
\begin{equation}
X \equiv \frac{|\mathbf{B}|}{\kappa_T k_B T \sum_{\alpha=1}^M \sum_{\beta=1}^M |\mathbf{B}|_{\alpha\beta}} - 1.  \label{eqn:X_def_KA}
\end{equation}
As for single-component systems, the isothermal compressibility for
this multicomponent system is obtained by computing the virial pressure
response to an incremental change in volume using Eq.~(\ref{eqn:kappa_T}).

For the Kob-Andersen system, we use a change of volume $\Delta V / V = 0.2\%$
and the pressure is sampled from $t = 20$ to $t = 40$, where $t = 0$ denotes
the time at which the quenching is halted and the system is compressed.
As can be seen in Fig.~\ref{fig:KA_X}, $X$ is zero\cite{footnote:zeroX}
for $T > 2 T_g$.
Similarly to the phenomenon observed in the case of the Z2 Dzugutov system
(see Fig.~\ref{fig:Z2_X}), $X$ increases up to a value of a value of
$\sim 0.15$ at $T = T_g$. The asymptotic behavior of $X$ for $T < T_g$ is
again the consequence of the system inability to relax in a time comparable
to the cooling schedule time per decade $\tau_{10}$.

\section{Conclusions and Discussion \label{sec:conclusion}}

We have demonstrated here that the static structural length scales $\xi_c$ and $\xi_C$
are able to distinguish subtle structural differences between glassy and liquid states,
which extends the analogous results for metastable hard spheres \cite{Ho12}
to atomic thermal systems.
Since these length scales are based on the volume
integral of the direct correlation function $c(r)$ and its generalization
$C(r)$, respectively, their growth as a liquid is cooled past its glass 
transition is a sign of the presence of long-range correlations in the glassy 
state that are not present in liquids. Additionally, the continuing increase
of $\xi_c$ and $\xi_C$ past the glass transition indicates that,
while particles primarily undergo sequences of local rearrangements, 
the glass may still
exhibit order on a significantly larger length scale as the system continues
to cool.
Our results using two-dimensional Kob-Andersen binary mixtures and
three-dimensional Z2 Dzugutov single-component systems, as well as the
previous results for MRJ packings as evidence,
we postulate that these length scales are relevant in various glasses.
This includes not only atomic systems possessing pair potentials with
steep repulsions and short-range attractions, but network
glasses as well. For example, in a recent computational study,\cite{Henja2012} 
which is supported by recent experimental 
results,\cite{LXWMRTS2012a} it was shown that realistic models of
amorphous silicon can be constructed to
be nearly hyperuniform, which implies that such glassy tetrahedrally-coordinated networks
are characterized by a large static length scale $\xi_c$
We also have shown that the nonequilibrium index $X$ is positively correlated with
a characteristic relaxation time scale, since they both increase as a system
is supercooled.

An interesting issue concerns the explication of
the underlying geometrical reasons for the
negative algebraic tail in the pair correlation function,\cite{DST2005a}
which also has been observed in hard-sphere systems.\cite{Ho12}
(The former is exhibited for large but bounded pair distances,
while the latter is valid asymptotically as $r \rightarrow \infty$.)
The local geometric diversity of particle arrangements in an amorphous solid 
medium inevitably creates short-range density fluctuations.  In particular, 
this is true for the nearly hyperuniform cases examined in this study.  Without being
too specific, one can formally divide a ``jammed'' particle configuration
into two equal subsets containing particles experiencing either
lower or higher local densities than the overall system average.  The fact
that the pair correlation functions display negative algebraic
tails with increasing separation $r$ has basic implications for the relative
spatial distributions of these low and high local density particles.  In
particular, it indicates that large numbers of either particle type cannot fit
together to form arbitrarily large clusters that dominantly exclude the other
particle types.  Instead, their spatial patterns evidently involve
interpenetrating percolating networks in three dimensions and highly
non-convex clusters in two dimensions.  The detailed statistical geometric
description of these patterns and why they generate algebraic pair correlation
function tails constitutes an important area for future investigation.

The quantity $X$ introduced earlier in Eq.~(\ref{eqn:X_def})
as a measure of deviation from thermal
equilibrium can be usefully interpreted in terms of system occupancy on the
many-body potential energy landscape.\cite{Stillinger1998}
Specifically, this focuses on the
comparative behaviors of isothermal compressibility at high-temperature
thermal equilibrium in the liquid phase as opposed to the measured isothermal
compressibility in the non-equilibrium glass phase in the $T \rightarrow 0$
limit.  In the former case, an incremental pressure change and accompanying
volume change will include shifts in occupancy probabilities for the separate
basins that tile the landscape; these shifts involve interbasin local particle
rearrangements that act to enhance the volume change induced by the pressure
perturbation.  In contrast, at very low temperatures, the system is trapped in
its initial basin; intrabasin vibrational motions have insufficient
amplitude to allow the system to take advantage of the previous kinds of local
particle rearrangements.  The resulting absence of enhanced volume change due
to those interbasin transitions reduces isothermal compressibility, causing $X$
to increase above zero.

\section*{Acknowledgements}
\vspace{-0.2in}
We are deeply grateful to Adam Hopkins for providing insights concerning
the generation of the supercooled states and the computation of the
compressibility.
We thanks Steven Atkinson for a critical reading of the manuscript.
This work was
supported by the Office of Basic Energy Science, Division
of Materials Science and Engineering under Award No. DEFG02-
04-ER46108.
S.~T. gratefully acknowledges the support of a Simons Fellowship in 
Theoretical Physics, 
which has made his sabbatical leave this entire academic year possible.


\begin{thebibliography}{32}
\bibitem{footnote:rank-1} Using expression (\ref{eqn:S_using_coll_coords})
to write $\mathbf{S}(\mathbf{k})$ for a single configuration, we can express
$\mathbf{S}(\mathbf{k})$ as the product of a non-zero vector times its 
Hermitian transpose, the rank of which (equal to the number of
non-zero eigenvalues) is 1. 
Taking an ensemble average of Eq.~(\ref{eqn:S_using_coll_coords})
breaks this symmetry, since the sum of $M$ vectors multiplied by their
transpose has a rank of $M$ if the vectors are linearly independent.
\bibitem{footnote:zeroX} Note that we found a very small systematic
error due to a combination of the following factors: the thermostat,
relaxation during the compressibility computation,
and the finite difference method. To correct for this systematic
error, we added to the computed value of $X$ a constant such as to ensure that
$X = 0$ for the high-temperature liquid phase, for which the compressibility
relations (\ref{eqn:comp_relation}) and (\ref{eqn:kirkwood_compress}) are
satisfied. This constant is equal to $0.05$ for the Z2 Dzugutov potential,
and $0.015$ for the Kob-Andersen potential.
\end{thebibliography}
\end{document}